\documentclass[11pt]{article}
\usepackage{amsmath,amssymb,amsfonts,amsbsy}
\usepackage{epsfig,subfig}
\usepackage[usenames,dvipsnames]{color}
\usepackage[authoryear,round]{natbib}
\usepackage{algorithm}
\usepackage[letterpaper, margin=1in]{geometry}
\usepackage{arydshln}
\usepackage{booktabs}
\usepackage{multirow}
\usepackage{epstopdf}
\usepackage{bbm}

\newtheorem{theorem}{Theorem}

\title{A Note on Automatic Data Transformation}
\author{Qing Feng, Jan Hannig and J.S.Marron}
\author{
	{\sc Qing Feng, Jan Hannig and J.S.Marron}\\
	Department of Statistics and Operations Research\\
	The University of North Carolina at Chapel Hill\\}
\date{\vspace{-5ex}}

\date{\vspace{-5ex}}

\begin{document}
	\maketitle
	
	\begin{abstract}
		Modern data analysis frequently involves variables with highly non-Gaussian marginal distributions. However, commonly used analysis methods are most effective with roughly Gaussian data. This paper introduces an automatic transformation that improves the closeness of distributions to normality. For each variable, a new family of parametrizations of the shifted logarithm transformation is proposed, which is unique in treating the data as real-valued, and in allowing transformation for both left and right skewness within the single family. This also allows an automatic selection of the parameter value (which is crucial for high dimensional data with many variables to transform) by minimizing the Anderson-Darling test statistic of the transformed data. An application to image features extracted from melanoma microscopy slides demonstrate the utility of the proposed transformation in addressing data with excessive skewness, heteroscedasticity and influential observations.
	\end{abstract}

	\textbf{keywords}: Automatic Transformation, \ Shifted Logarithm Transformation,  \ Anderson-Darling Test Statistics, \ Heteroscedasticity.

\section{Introduction}
\label{sec: intro}
Technological developments have led to methods for generating complex data objects such as DNA chip data and digital images of tumors. These new types of data objects frequently strongly violate the approximate normality assumption which is commonly made in statistical techniques. Therefore, an appropriate data transformation can be very useful for improving the closeness of the data distribution to normality.

Many transformation techniques have been proposed. \citet{sakia1992box} provided a comprehensive review of the Box-Cox~\citep{box1964analysis} and related transformations. Various methods have been developed for selecting the transformation parameters, including the maximum likelihood method \citep{box1964analysis}, robust adaptive method~\citep{carroll1980robust}, Kullback-Leibler information based method~\citep{hernandez1980large},  and Kendall's rank correlation based method~\citep{han1987non}. 

A commonly used member of the Box-Cox family is the logarithm transformation, which is useful for tackling data sets generated by a multiplicative process. Furthermore, the logarithm transformation can stabilize the asymptotic variance of data. One important application is to transform some types of microarray data. A shift parameter was further introduced to make the logarithm transformations more flexible and useful. See Section 3 of~\citet{Yang1995} for a good overview of the shifted logarithm transformation. The parameterizations of the shift parameter strongly depend on knowledge of the data e.g. data range, data distribution, so user intervention is usually required. However, modern high-output data sets usually have a very large number of variables, i.e. features, so there is a strong need to automate the selection of shift parameter, which is an important contribution of this paper. 

We propose a new automatic data transformation scheme for making various types of marginal distributions close to being normally distributed. In particular, we aim at addressing certain types of departures from normality e.g. strong skewness. Our proposed method focuses on the family of shifted logarithm transformations and introduces a new parametrization which treats the data as lying on the entire real line. Besides, our parametrization makes the selection of shift tuning parameter independent of data magnitude which is an advantage for automation. This algorithm is designed to automatically select a parameter value such that the transformed data has the smallest \emph{Anderson-Darling test statistic}. Furthermore, this transformation scheme includes a winsorisation of influential observations based on the extreme value theorem.

\subsection{Data Example}
\begin{figure}[ht]
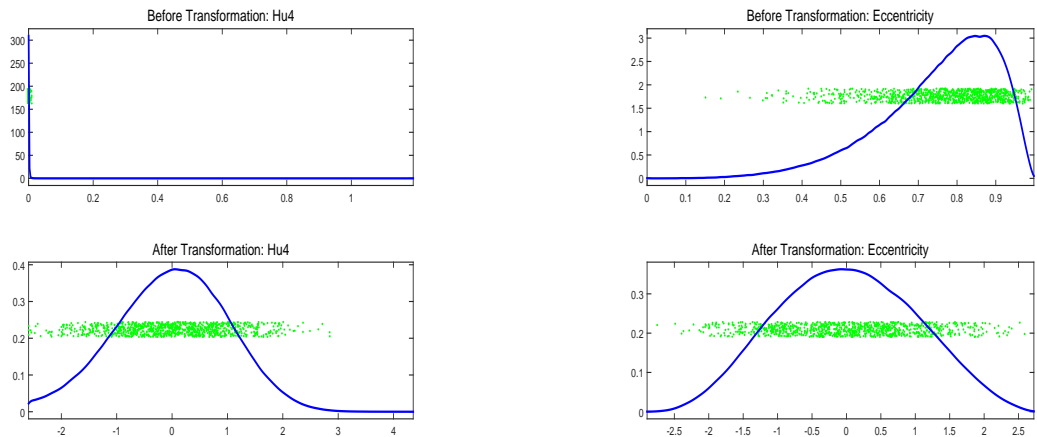

	\begin{minipage}[b]{0.5\linewidth}
		\centering
		\includegraphics[width=0.65\textwidth, height = 0.7\textwidth]{Hu4.eps}
	\end{minipage}
	\begin{minipage}[b]{0.5\linewidth}
		\centering
		\includegraphics[width=0.65\textwidth, height = 0.7\textwidth]{Eccentricity.eps}
	\end{minipage}
	\caption{Comparison of the KDE-plots of two image feature vectors before (top row) and after (bottom row) transformation. This shows that the transformed distributions are much closer to Gaussian for data with both positive (Hu4, left column)  and negative (Eccentricity, right column) skewness.}
	\label{fig:trans:kde3vs8}
\end{figure}

A motivating data example is digital image analysis in a study of mutant types of melanocytic lesions~\citep{miedema2012image}. Many of the raw features extracted from digital images contain excessive skewness. For example, the marginal distributions of two of the image features are visualized by the \emph{kernel density estimated plots} (KDE plots) in the top row of Figure~\ref{fig:trans:kde3vs8}. The blue curves are the Gaussian kernel density estimate i.e. smoothed histograms, using Sheather-Jones plug-in bandwidths. (See Chapter 3 of \citet{wand1994kernel} for the comparison of bandwidth selection methods.) The green dots are jitter plots of the data. Each symbol is a data point whose horizontal coordinate is the value and vertical coordinate is based on data ordering for visual separation. As can be seen, these distributions are highly skewed. For such data sets with substantial skewness, an analysis based on a Gaussian assumption would tend to generate poor results. The bottom plots of Figure~\ref{fig:trans:kde3vs8} display the KDE plots of each feature vector after our automatic transformation. The kernel density estimates (blue curves) are approximately symmetric.

\begin{figure*}[ht!]
	\begin{minipage}[b]{0.5\linewidth}
		\centering
		\includegraphics[scale=0.35]{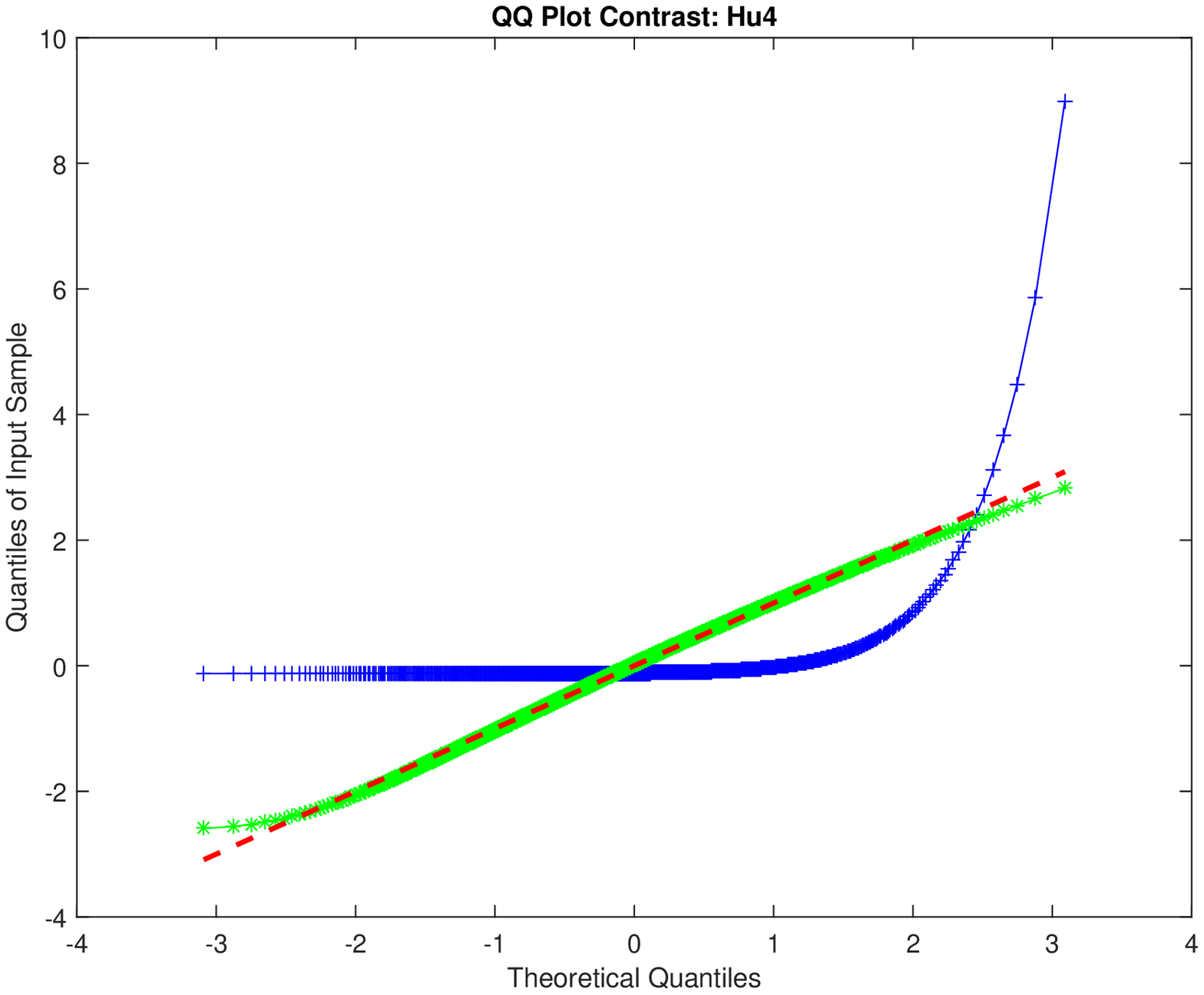}
	\end{minipage}
	\begin{minipage}[b]{0.5\linewidth}
		\centering
		\includegraphics[scale=0.35]{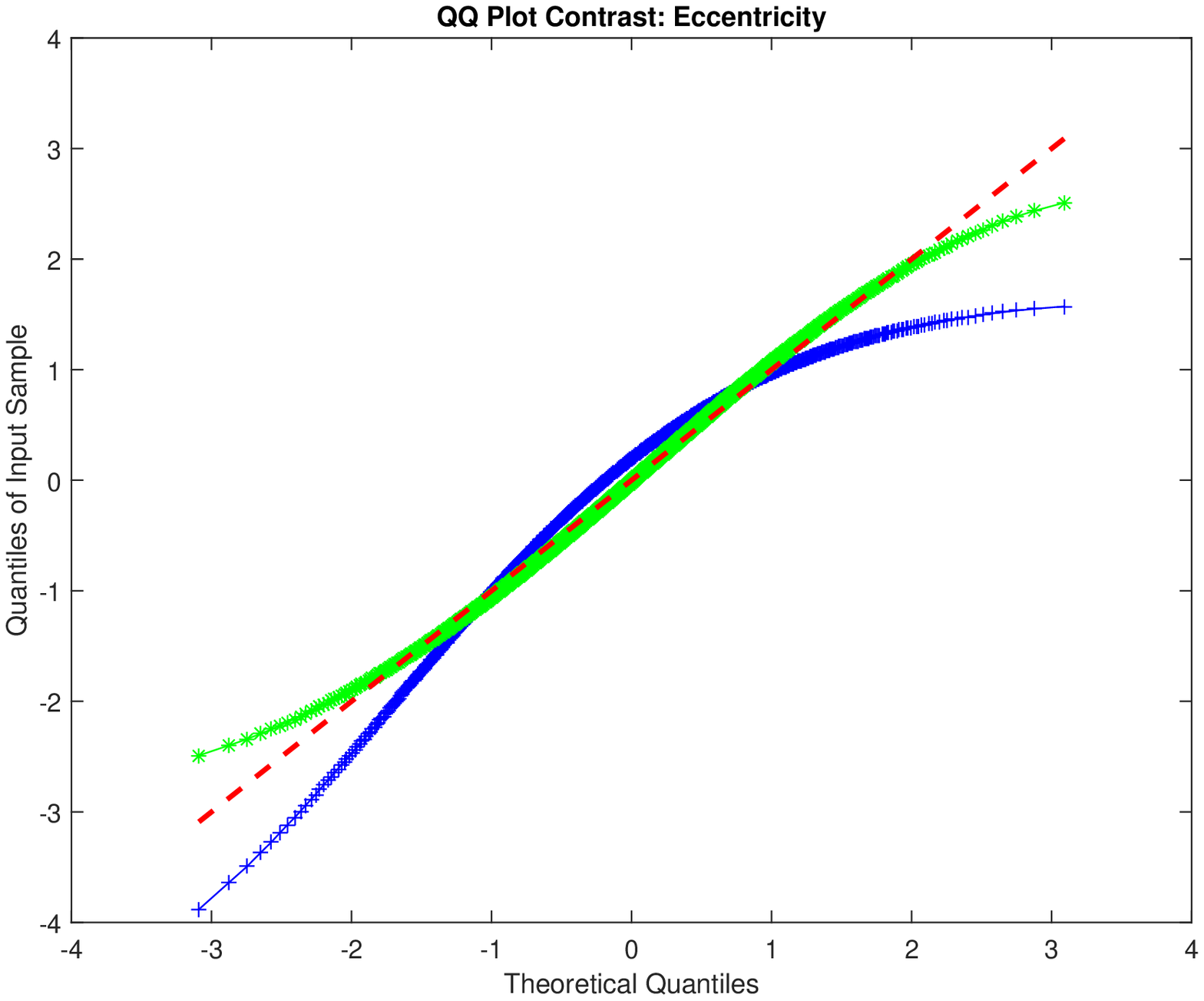}
	\end{minipage}
	\caption{The QQ-plots of Hu4 (left) and Eccentricity (right). The comparison between before (blue plus signs) and after (green stars) indicates a major overall improvement in closeness to normality made by transformation.}
	\label{fig:QQ}
\end{figure*}

The Q-Q Plot in Figure~\ref{fig:QQ} gives a more precise measure of closeness to the standard normal distribution. The left panel shows the Q-Q plots for Hu4 applied with standardization only (blue plus signs)  and for Hu4 after automatic transformation (green stars). The symbols are the quantiles of 1000 randomly selected data points against the theoretical quantiles of the standard normal distribution. For comparison, we also show the 45 degree red dashed line. The blue plus signs clearly depart from this line, while the green stars approximately lie on the line. This contrast suggests a dramatic improvement in normality by our automatic transformation of Hu4. A similar improvement in normality of Eccentricity is also shown in the right panel. Although there are slight departures at each tail of the transformed data, an overall improvement can be seen as the majority of the quantiles approach the theoretical quantiles of the standard normal distribution.

\begin{figure*}[ht!]
	\begin{center}
		\includegraphics[trim={0 0 0 1.4cm},clip, width=0.55\textwidth, height = 0.35\textwidth]{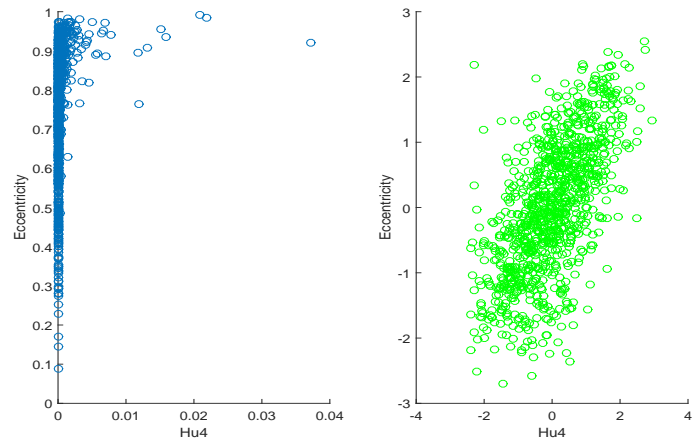}
	\end{center}
	\caption{Comparison of the scatter plots, showing the joint distributions from Figure~\ref{fig:trans:kde3vs8}, before (left) and after (right) transformation. Relationship after transformation is much closer to linear.}
	\label{fig:trans:scatterintro}
\end{figure*}

Even though our transformation acts only on the marginal distributions, it often results in major improvement of the joint distribution of the features. In Figure~\ref{fig:trans:scatterintro}, the scatter plot on the left shows a strong non-linear relationship between the Hu4 and Eccentricity that were studied in Figures~\ref{fig:trans:kde3vs8} and~\ref{fig:QQ}. After transformation, the scatter plot on the right shows a bivariate Gaussian relationship which is much more amenable to analysis using standard statistical tools.

\section{Methodology}
\label{sec: method}
In this section, a novel automatic data transformation scheme is proposed for general data sets to achieve approximate normality. For any given data set, the transformation works feature by feature. In other words, for a data matrix with columns considered as data objects and rows considered as features, the transformation is applied to each row.

The transformation scheme consists of three components: a family of shifted logarithm transformation functions indexed by a parameter $\beta$, standardization with an option for winsorisation of extreme observations and an evaluation of the transformation with a given parameter value. The key steps will be introduced in the following subsections.

The transformation scheme is a grid search based on three components to determine the optimal value of $\beta$ for each feature, which is outlined as
\begin{itemize}
	\item \emph{Initialization:} Construct a grid of parameter values $\beta = \{\beta_{k}, k=1, \cdots, n\}$
	\item \emph{Step 1:} Apply the transformation function to the feature vector for each parameter value $\beta_{k}$.
	\item \emph{Step 2:} Standardize the transformed feature vector and winsorise any existing extreme observations. Re-standardize the feature vector if winsorisation has been done.
	\item \emph{Step 3:} Calculate the Anderson-Darling test statistic.
\end{itemize}
Lastly, select $\beta$ to minimize the Anderson-Darling test statistic for normality.

\subsection{Transformation Function}
A new parametrization of the family of shifted logarithm functions, $\{\phi_{\beta}, \beta \in \mathbb{R}\}$, is proposed for addressing both left and right skewness in features. For a feature vector $X$ = $(X_{1},X_{2},\cdots, X_{n})$, the sample skewness is 
$
g(X) = \frac{\frac{1}{n}\sum_{i=1}^{n}(X_i-\bar{X})^3}{(\frac{1}{n}\sum_{i=1}^{n}(X_i-\bar{X})^2)^{\frac{3}{2}}}
$
where $\bar{X}$ is the sample mean of the vector.

As convex transformation functions tend to increase the skewness of data while concave transformations reduce it \citep{van1964convex}, the transformation functions are chosen to be concave for $g(X)>0$ and convex for $g(X)<0$.
As logarithm functions are concave, the transformation function can be a logarithm for the case $g(X)>0$ i.e. $\log(X_{i})$. While for the other case $g(X)<0$, the transformation function should be made convex by inserting negative signs within and before a logarithm function i.e. $-\log(-X_{i})$.

Since logarithm functions require positive inputs, it is important to modify the functions for both cases to be valid for any $X_{i}$. For example, in the case $g(X)>0$, this concern can be resolved by subtracting the minimal value of the feature vectors from $X_{i}$ and adding a positive shift parameter $\alpha$. That is,
\begin{equation}
\log(X_{i}-min(X_{1}, X_{2}, \cdots, X_{n}) + \alpha),
\end{equation}
Similarly for the negative skewness $g(X)<0$, the function is
\begin{equation}
-\log(max(X_{1}, X_{2}, \cdots, X_{n})-X_{i}+\alpha),
\end{equation}

The shift parameter $\alpha$ is further parameterized in terms of the multiples of the range of the feature vectors i.e. $R$ = $max(X_{1},X_{2},\cdots, X_{n})$ - $min(X_{1},X_{2},\cdots, X_{n})$. This makes the selection of parameter values independent of the data magnitude. In particular, set
\begin{equation}
\alpha = |\frac{1}{\beta}|R.
\end{equation}
By tuning the value of $\beta$, the effect of the transformation varies. In particular, the transformation together with standardization is equivalent to standardization only, when the parameter $\beta$ approaches $0$. In order to make the resulting transformation function $\phi_{\beta}(X_{i})$ continuous over $\beta \in \mathbb{R}$, we define our transformation to be standardization only for $\beta = 0$. 

Incorporating all these elements, the formal representation of the family of transformation functions is
\begin{equation}
\label{equ:trans:function}
\phi_{\beta}(X_{i}) = \left\{ \begin{array}{rl}
log(X_{i}-min(X_{1}, X_{2}, \cdots, X_{n})+|\frac{1}{\beta}|R), &\mbox{$\beta>0$}\\
-log(max(X_{1}, X_{2}, \cdots, X_{n})-X_{i}+|\frac{1}{\beta}|R), &\mbox{$\beta<0$}\\
\end{array} \right.
\end{equation}
in which $\beta \in \mathbb{R}$, $R$ and $g(X)$ are as defined above.

\subsection{Standardization and Winsorisation}
Standardization is applied to the transformed feature vector i.e.$[\phi_{\beta}(X_1), \cdots, \phi_{\beta}(X_n)]$, by subtracting its median and dividing by the mean absolute deviation from the median\footnote{If the mean absolute deviation is zero, return a vector of zeros.}. Denote the vector after standardization as $X^{\dag}$. A winsorisation of $X^{\dag}$ at an appropriate threshold is further applied to reduce the influence of extreme observations.

\subsubsection{Winsorisation}
Extreme value theory provides reasonable choices of thresholds for winsorisation. A fundamental result of that area is the \emph{Three Types Theorem}. See \citet{leadbetter2011extremes} for detailed discussion.

\begin{theorem}[The Extremal Types Theorem]
	Suppose $X=(X_{1},X_{2},\cdots, X_{n})$ are independent, identically distributed standard normal random variables, there exist real constants $a_{n}>0$ and $b_n$ such that
	\begin{equation}
	P(\frac{M_n-b_n}{a_n} \leq x) \rightarrow G(x)
	\end{equation}
	where $G(x)$ is the cumulative distribution function of the standard Gumbel distribution i.e. $G(x)=e^{-e^{-x}}$ and
	\begin{equation}
	b_n=(2\log n)^{\frac{1}{2}}-\frac{\log\log n+\log(4\pi)}{(2\log n)^{\frac{1}{2}}}
	\end{equation}
	\begin{equation}
	a_n=(2\log n)^{-\frac{1}{2}}
	\end{equation}
\end{theorem}
From this extreme value theory, the threshold of the standardized vector $X^{\dag}$ is computed based on the 95th percentile of the standard Gumbel distribution ($p_{95}$), that is
\begin{equation}
\label{equ:trans:threshold}
L=p_{95}a_n+b_n.
\end{equation}
When the absolute value of the element in $X^{\dag}$ is greater than $L$ i.e. $|X_i^{\dag}|>L$, the element value is winsorized (i.e pulled back) to the value $sign(X_i^{\dag})L$. After the winsorisation, the feature vector will be standardized again, by substracting the sample mean and dividing by the standard deviation.

\subsection{Evaluation}
The evaluation of the stated transformation procedure is based on measuring the distance between the empirical distribution function (EDF) of the transformed data and the cumulative distribution function (CDF) of the standard normal. Commonly used EDF statistics are the Kolmogorov-Smirnov test statistic, the Cram\'{e}r-von Mises test statistic, the Watson statistic and the Anderson Darling test statistic. \cite{stephens1974edf} conducted power studies of these statistics under different specifications of hypothesized distributions. Based on this study, the Anderson Darling test statistic is considered as powerful for detecting most common departures from normality. Therefore, that is used here as the criterion for evaluation.
A computable form of the Anderson Darling test statistic is defined in terms of the order statistics i.e.
\begin{equation}
A^2 = -n - \sum_{i=1}^{n}\frac{2i-1}{n}[\log \Phi(X_{(i)})+\log (1-\Phi(X_{(n+1-i)}))]
\end{equation}
Larger values of this indicate stronger departures from Gaussianity. Thus, by searching for a parameter value minimizing this statistic, an optimal transformation for improving the closeness of the distributions of features to normality is obtained.

\bibliographystyle{plainnat}
\bibliography{autotransreference}

\end{document}